\documentclass[12pt,aps,floats,showpacs,amssymb,tightenlines]{revtex4}
\usepackage{amsmath}
\usepackage{amsfonts}
\usepackage{amscd}
\usepackage{epsfig}
\usepackage{amssymb}
\usepackage{tabularx}
\renewcommand{\a}{\alpha}

\begin{document}

\title{Static black hole solutions with a self interacting conformally coupled scalar field}
\author{Gustavo Dotti}  \email{gdotti@famaf.unc.edu.ar} \author{Reinaldo
J. Gleiser } \email{gleiser@fis.uncor.edu} \affiliation{Facultad de
Matem\'atica, Astronom\'{\i}a y F\'{\i}sica, Universidad Nacional de
C\'ordoba, Ciudad Universitaria, (5000) C\'ordoba, Argentina}
\author{Cristi\'an Mart\'\i nez}\email{martinez@cecs.cl} \affiliation{Centro
de Estudios Cient\'\i ficos (CECS), Casilla 1469, Valdivia, Chile\\and Centro de Ingenier\'{\i}a de la Innovaci\'on del CECS (CIN), Valdivia,
Chile}

\begin{abstract} We study static, spherically symmetric black hole solutions
of the Einstein equations with a positive cosmological constant and
a conformally coupled self interacting scalar field. Exact solutions
for this model found by Mart\'\i nez, Troncoso, and Zanelli, (MTZ),
were subsequently shown to be unstable under linear gravitational perturbations,
with modes that diverge arbitrarily fast. We find that the moduli
space of static, spherically symmetric solutions that have  a
regular horizon -and satisfy the weak and dominant energy conditions
outside the horizon- is a singular  subset of a two dimensional
space parameterized by the horizon radius and the value of the
scalar field at the horizon. The singularity of this space of
solutions provides an explanation for the instability of the MTZ
spacetimes, and leads to the conclusion that, if we include
stability as a criterion, there are no physically acceptable black
hole solutions for this system that contain a cosmological horizon
in the exterior of its event horizon.
\end{abstract}

\maketitle

\noindent

\section{Introduction} \label{introduccion}

When one considers possible fields interacting with a black hole,
the simplest source of matter that one could naively take into account
corresponds to a single real scalar field. However, when this field
is minimally coupled and the spacetime is asymptotically flat, the
so called no-hair conjecture
\cite{Ruffini-Wheeler,Bekenstein:1971hc,Teitelboim-No-hair}
indicates that this class of black hole does not exist. Much effort
have been focused on this problem and recent works dealing with this
issue can be found in \cite{recentwork}. Nonetheless, this
conjecture can be circumvented in different ways as we show below.

A black hole solution, where the scalar field is
\textit{conformally} coupled, i. e. when the corresponding
stress-energy tensor is traceless, was found in
\cite{Martinez:1996gn}. In this three-dimensional black hole, the
scalar field is regular everywhere and the spacetime is
asymptotically anti-de Sitter because a negative
cosmological constant is included.  This black hole solution can be
extended by considering a conformal self-interacting potential. This
was done in
\cite{Henneaux:2002wm}, where  exact black hole
solutions are found for  a minimally coupled scalar field and a one parameter family of
potentials. A previous
four-dimensional and asymptotically flat black hole \cite{BBMB} was
reported back in the 70's, but the scalar field diverges at the horizon.
The presence of a cosmological constant allows to find
exact four-dimensional black hole solutions, where the scalar field
is regular on and outside  the event horizon
\cite{Martinez:2002ru,Martinez:2004nb,Martinez:2005di,Martinez:2006an}.
Numerical black hole solutions can  also be found in four
\cite{Torii:2001pg,Winstanley:2002jt,Hertog:2004bb,Winstanley:2005fu,Radu:2005bp}
and five dimensions \cite{Hertog:2004dr}. Further exact solutions
in the context of low energy string theory were found in \cite{Zloshchastiev:2004ny}.

Some interesting aspects of these black hole solutions are studied  in
 \cite{several}. In particular, the analysis of
stability against linear perturbations for the de Sitter conformally
dressed black hole \cite{Martinez:2002ru} done in \cite{HTWY},  is
relevant for the discussion presented here.\\

In this work, we study the space of static, spherically symmetric
solutions of the Einstein equations with a positive cosmological
constant and a conformally coupled self-interacting scalar field
(MTZ model). Conformally coupled scalar fields
in General Relativity have been  used to model quantum effects in semiclassical theories \cite{qc}.
This model has a well posed initial value formulation \cite{wp}, and  was shown to
reproduce better -than the minimally coupled scalar field-
the local propagation properties of Klein Gordon fields on Minkowski
spacetime \cite{Sonego}. Our interest, however, comes from the fact that this model
 allows
non trivial static black holes solutions \cite{Martinez:2002ru}.
These solutions  belong to a restricted class (eq. (\ref{MTZ5}) below) of spherically
symmetric static spacetimes, and are given  in
 equations  (\ref{1MTZ}) (\ref{1MTZs}) (solution MTZ1)
and (\ref{exa1}) (solution MTZ2). Note that a generic  spherically
symmetric static spacetime metric  admits the local form
(\ref{met1}). In this work we address the following question: Are
there other static, spherically symmetric black hole solutions for
the MTZ model, satisfying the dominant and strong energy condition
between the event and cosmological horizon, besides MTZ1 and MTZ2?
 Using a combination of analytical and numerical methods we conclude that
the answer to this question is negative.\\

The paper is organized as follows: in the next section we review the
MTZ model and skecth the derivation of the MTZ1 and MTZ2 solutions.
We also prove that MTZ1 is unstable under spherically symmetric gravitational perturbations.
MTZ2 had already been found to be unstable under spherically symmetric gravitational perturbations,
this being our original motivation to study the space of spherically symmetric static solutions
of the MTZ system. This is done in
 Section \ref{metric}, where the full set of Einstein and scalar field equations is reduced to
 a second order ODE system.
  In Section
\ref{reghor} we analyze the restrictions that the existence of a
regular event horizon impose on the solutions, if we also require
that the energy-momentum tensor satisfies appropriate energy
conditions. Acceptable local solutions are found to be
parameterizable with the horizon radius $r_0$ and the value of the
scalar field at the horizon, $a_0 := \phi(r_0)$. The subset of
allowed values is displayed in Fig. \ref{modspace}. To address the
issue of the global behavior of these local solutions, the field
equations were numerically integrated away from the horizon. Some
illustrative examples are presented in Section \ref{numerico}, where
the different behaviors as we move away from the event horizon are
shown. We find that solutions that satisfy the energy conditions near the event horizon
contain, in general, a coordinate singularity for some finite $r$
outside the event horizon. We show that the isotropy spheres reach a maximum radius $r$
at this point and contract as the proper distance from the horizon further increases.
This explains  why $r$ is not an appropriate coordinate in this region.
We provide  an appropriate coordinate extension in Section \ref{coord1}. A
numerical integration beyond the coordinate singularity, described
in Section \ref{numerico1}, suggests   that, generically, the metrics
contain a curvature singularity (the energy density diverges) at
some finite proper distance from the extension point.

Mostly for completeness we include in Section \ref{eviol} an
analysis of the metrics that violate the energy conditions. A
compilation of the main results, together with some final comments
and our conclusions are given in Section \ref{conclus}

\section{The MTZ model} \label{intro}

In the MTZ model \cite{Martinez:2002ru} the action for gravity conformally
coupled to a scalar field $\phi$ with a quartic self-interaction
potential and an electromagnetic field $F_{\mu\nu}$ is given by,
\begin{eqnarray}
\label{MTZ1}
  S &=& \frac{1}{2}\int{d^4x \sqrt{-g}\left[R-2\Lambda-g^{\mu\nu} \partial_{\mu}\phi
  \partial_{\nu}\phi -\frac{1}{6} R \phi^2 \right.} \nonumber \\
  & & \left. -2 \alpha \phi^4 -\frac{1}{8\pi} F^{\mu\nu}F_{\mu\nu}
   \right], \end{eqnarray}
where $\alpha$ is a coupling constant. Variation of this  action
with respect to the metric, scalar field and Maxwell potential gives the
following set of Euler Lagrange equations:

 \begin{subequations} \label{el}
      \begin{eqnarray}
         G_{\mu\nu}+\Lambda g_{\mu\nu} &=&
T^{\phi}_{\mu\nu} + T^{EM}_{\mu\nu} \label{el-a} \\
\Box \phi  -\frac{1}{6} R \phi -4
\alpha \phi^3 &=&   0 \label{el-b} \\
\nabla^{\mu} F_{\mu\nu} &=& 0 , \label{el-c} 
       \end{eqnarray}
where the stress energy tensors are
\begin{equation}  \label{el-d}
 T^{\phi}_{\mu\nu} =
\partial_{\mu} \phi
\partial_{\nu} \phi -\frac{1}{2} g_{\mu\nu} g^{\alpha \beta}
\partial_{\alpha} \phi \partial_{\beta}
\phi  +\frac{1}{6} \left[g_{\mu\nu}  \Box -\nabla_{\mu}\nabla_{\nu}
 + G_{\mu\nu} \right] \phi^2 -\alpha g_{\mu\nu} \phi^4
 \end{equation}
 and
 \begin{equation} \label{el-e}
T^{EM}_{\mu\nu}  = \frac{1}{4\pi}\left(g^{\alpha\beta}
F_{\mu\alpha} F_{\nu\beta}-\frac{1}{4} g_{\mu\nu}F_{\alpha \beta}
F^{\alpha\beta} \right) .
 \end{equation}
\end{subequations}
Under conformal transformations $\phi \to \Omega^{-1} \phi$,
$F_{\mu\nu} \to F_{\mu\nu}$, $g_{\mu\nu} \to \Omega^{ 2}
g_{\mu\nu}$,  equations (\ref{el-b})-(\ref{el-c}) are invariant, and
$T^{\phi}_{\mu\nu} \to \Omega^{-2} T^{\phi}_{\mu\nu},\;
T^{EM}_{\mu\nu} \to \Omega^{-2} T^{EM}_{\mu\nu}.$ This is the
motivation behind the choice of the non minimal coupling and quartic
self interaction of the scalar field.

Note that the trace of $T^{\phi}_{\mu\nu}$ vanishes on shell:
\begin{equation} \label{tracefi}
T^{\phi} := T^{\phi}_{\mu\nu} g^{\mu \nu} = \phi \left[ \Box \phi -
\frac{R}{6} \phi -4 \alpha  \phi^3 \right]
\end{equation}
whereas $T^{EM} := T^{EM}_{\mu\nu} g^{\mu \nu}$ vanishes identically.
Thus, taking the trace of Eq. (\ref{el-a}) gives
\begin{equation} \label{se}
R = 4 \Lambda
\end{equation}
{\em We should stress here that (\ref{se}) does not follow from (\ref{el-a}) alone, but from the
system (\ref{el-a})-(\ref{el-b})-(\ref{el-d})-(\ref{el-e}).}\\
It is  interesting to comment on those solutions of the field equations (\ref{el})
for which $\phi \equiv  \phi_o$,  $\phi_o \neq 0$ a  constant
(we are not interested in the pure Einstein-Maxwell case
 $\phi \equiv 0$). In this case, the system (\ref{el})  reduces to:
 \begin{subequations} \label{els}
      \begin{eqnarray}
\left( 1 - \frac{\phi_o{}^2}{6} \right) G_{\mu\nu}+ ( \Lambda + \alpha \phi_o{}^4 )  g_{\mu\nu} &=&
T_{\mu \nu}^{EM} \label{els-a} \\
 R  + 24
\alpha \phi_o{}^2  &=&   0 \label{els-b} \\
\nabla^{\mu} F_{\mu\nu} &=&0. \label{els-c} 
       \end{eqnarray}
\end{subequations}
Taking the trace of 
  (\ref{els-a}) and using (\ref{els-b}) gives
\begin{equation} \label{cf}
\phi_o{}^2 =  - \frac{\Lambda}{6 \alpha}.
\end{equation}
Thus (\ref{els-a}) takes a simple form:
\begin{equation}
\label{MTZ2.2}
\left( 1 + \frac{\Lambda}{36 \alpha} \right) \; \left[ G_{\mu\nu}+\Lambda g_{\mu\nu} \right] =
T^{EM}_{\mu \nu}, 
\end{equation}
and (\ref{els-b}) gives again (\ref{se}).
Note that these are Einstein-Maxwell  equations with an effective Newton's constant $G_{eff} =
\left( 1 + \frac{\Lambda}{36 \alpha} \right)^{-1} G$ \cite{Martinez:2002ru}, thus
the case where $\left( 1 + \frac{\Lambda}{36 \alpha} \right) < 0$
(negative $G_{eff}$) is somewhat pathological because it is
equivalent to having repulsive gravitational forces
\cite{Martinez:2002ru}.
\\
The theory with a coupling constant $\a$  tuned with the cosmological constant as
\begin{equation} \label{st}
\a = - \frac{\Lambda}{36} ,
\end{equation}
is particularly interesting, since it seems to admit a wider set of solutions.
We will call these theories {\em special} from now on, and call $\alpha \neq  -\Lambda/36$ theories
{\em generic}.
For special theories and  constant scalar field configurations, $\phi_o{}^2=6$ and the field 
equations (\ref{els}) 
 become
\begin{subequations} \label{elss}
      \begin{eqnarray}
0 &=& T_{\mu \nu}^{EM}
 \label{elss-a} \\
 R  - 4 \Lambda   &=&   0 \label{elss-b} \\
\nabla^{\mu} F_{\mu\nu} &=&0 , \label{elss-c} 
       \end{eqnarray}
\end{subequations}
Note that the Euler-Lagrange equation for the metric, eq. (\ref{elss-a}), gives
no information about the metric, but implies $F_{\mu \nu}=0$.
{\em This does not mean that the gravitational field is unconstrained}, as one
might first be lead to think, since the
Euler-Lagrange equation for the {\em scalar field}
forces $R=4 \Lambda$ in this case, so we do get an equation for the metric (note in pass
the $R=$constant follows just from the scalar field equation (\ref{el-b}) when $\phi=$ constant). \\

In this paper we will consider only the  case
$F_{\mu\nu}=0$, and will explore the space of static, spherically
symmetric solutions:
\begin{equation}\label{met1}
ds^2 = -N_2(r)\; dt^2 + N_1(r) \; dr^2 + r^2 \; d\Omega^2, \;\; \phi = \phi(r).
\end{equation}
Since all solutions of the field equations satisfy
(\ref{se}), we will oftentimes replace
 eq. (\ref{el-b}) with the much simpler equation
\begin{equation}
\label{MTZ3.1}
\Box \phi  -\frac{2}{3} \Lambda \phi -4
\alpha \phi^3=0.
\end{equation}
We were naturally led to consider this problem from the linear
stability analysis in \cite{HTWY} of the exact solutions found in
\cite{Martinez:2002ru}. These exact solutions are all of the form
\begin{equation}\label{MTZ5}
ds^2 = -N(r) dt^2 + N(r)^{-1} dr^2 + r^2 d\Omega^2, \;\; \phi=\phi(r).
\end{equation}
The first one that we analyze,
which we call here solution MTZ1, has a constant scalar field (\ref{cf}).
For generic theories $N(r)$ is  obtained
by imposing on (\ref{MTZ5}) the condition $G_{\mu \nu} = - \Lambda g_{\mu \nu}$, which
follows from (\ref{els-a}), or more directly from (\ref{MTZ2.2}).
For the special theories (\ref{st}), as explained
above, the Euler-Lagrange equation for the metric is trivial, and the only constraint
on (\ref{MTZ5}) is $R=4 \Lambda$, and comes from the Euler-Lagrange equation (\ref{els-b}) for the scalar field.
 Since this
condition on the metric is less restrictive than the one  for generic theories, we get a wider set of
solutions for  special theories (two integration constants, $Q$ and $M$ below,  instead of one):
\begin{eqnarray} \label{1MTZ}
N(r) =    1 - \frac{2M}{r} -\frac{\Lambda}{3} r^2,  &\phi(r) = \sqrt{\frac{-\Lambda}{
 6 \alpha}}, &  \alpha \neq -\Lambda/36\\
 N(r) = 1 - \frac{2M}{r}+ \frac{Q}{r^2} -\frac{\Lambda}{3} r^2,  &\phi(r) = \sqrt{6}, &
 \alpha = -\Lambda/36.  \label{1MTZs}
\end{eqnarray}
In other words,  requiring
$R = 4 \Lambda$ to the metric (\ref{MTZ5}) gives $N(r)$ as
in (\ref{1MTZs}). Adding the extra condition $R_{\mu \nu} = \Lambda g_{\mu \nu}$
forces $Q=0$.
Note that (\ref{1MTZ}) is the  Schwarzchild-(A)dS  metric in the generic case,
Reissner-Nordstr\"om (A)dS for  special theories (with
$Q$ the ``source'' of a scalar field instead of the square of the electric
charge). \\
Since we are only interested in static black hole solutions with a physically acceptable
stress-energy-momentum tensor, we require that  the singularity at $r=0$  be hidden behind
an event horizon, and that $T^{\phi}_{\mu \nu}$   satisfies appropriate energy conditions
in the $N(r)>0$ region between the event and cosmological horizons, which we assume located at
$r>0$, using if necessary the invariance of the metric under $(r,M) \to (-r,-M)$).
MTZ1 has $T^{\phi}_{\mu \nu}= G_{\mu\nu}+\Lambda g_{\mu\nu} =0$ in the generic case.
 A straightforward calculation
shows that for the special theory (\ref{1MTZs})
\begin{equation} \label{ec1}
T_{\mu \nu} = \frac{Q}{r^4} \left( \hat t_{\mu} \hat t_{\nu}-
\hat r_{\mu} \hat r_{\nu}+\hat \theta_{\mu} \hat \theta_{\nu}+\hat \phi_{\mu} \hat \phi_{\nu}
\right)
\end{equation}
in the natural orthonormal basis $\hat t ^{\mu} = N^{-1/2} \; {\partial}_{t}, \;
\hat r^{\mu} = N^{1/2} \; {\partial}_{r},\; \hat \theta ^{\mu} = r^{-1}
{\partial}_{ \theta},\; \hat \phi ^{\mu} = (r \sin(\theta) )
 ^{-1} {\partial}_{\phi}$. Thus, the strong and dominant energy conditions are satisfied
 in both cases as long as $Q>0$.\\

The second type of $F_{\mu \nu}=0$ solution in \cite{Martinez:2002ru} for the
system (\ref{MTZ1}) and the Ansatz (\ref{MTZ5}), which we call MTZ2,
holds only for the special  theories $\alpha = -\Lambda/36$. The
metric is that of a Reissner-Nordstr\"om (A)dS black hole,
the mass being an integration constant that appears  both in the
metric and the scalar field:
\begin{eqnarray}
\label{exa1}
  \phi &=& \frac{\sqrt{6} M}{r-M} \nonumber \\
  N &=& \left(1-\frac{M}{r}\right)^2-\frac{\Lambda}{3} r^2.
\end{eqnarray}
To avoid naked singularities, we restrict to  the case $\Lambda >
0$, then $N \to - \infty $ as $r \to \infty$, $N \to \infty$ as $r
\to 0^+$, and the singularity at $r=0$  is not  naked
only if $N(r)$ has three positive roots. This can only happen if $0
< M < \sqrt{3}/(4\sqrt{\Lambda}) =: l/4$. In this case, the three
positive roots are
\begin{equation}
r_1 = \frac{l}{2} \left( -1 + \sqrt{1 + \frac{4M}{l}} \right)  < r_2 =
\frac{l}{2} \left( 1 - \sqrt{1 - \frac{4M}{l}}\right) < r_{3} =
\frac{l}{2} \left( 1 + \sqrt{1 - \frac{4M}{l}} \right)
\end{equation}
This solutions are  black holes on a cosmological background, with
an inner horizon $r_1$, a  regular event horizon $r_2$
and a cosmological horizon $r_3$ \cite{Martinez:2002ru}.\\

It will be useful for our discussion to review the derivation of the
MTZ metrics from the Ansatz (\ref{MTZ5}). Notice that the Einstein
plus scalar field equations imply that the functions $N(r)$ and
$\phi(r)$ must satisfy four equations, and therefore the set of
solutions is severely restricted. Assuming as stated that
$F_{\mu\nu}=0$, and the form (\ref{MTZ5}) for the metric, an
appropriate combination of the Einstein equations implies that
$\phi$ satisfies the equation,
\begin{equation}\label{mtza01}
\phi \frac{d^2\phi}{dr^2} - 2 \left(\frac{d\phi}{dr}\right)^2 = 0
\end{equation}
This admits the solution $\phi(r)=0$, leading to vacuum black holes
with $\Lambda \neq 0$, and also a  general solution of the form,
\begin{equation}\label{mtz02}
\phi(r)=\frac{1}{C_1 r+C_2}
\end{equation}
where $C_1$, and $C_2$ are constants. The two kinds
of solutions, MTZ1 and MTZ2, are obtained by choosing $C_1 = 0$ or $C_1 \neq 0$,
then solving the remaining field equations. {\em There is no other solution
to the field equations
of the form (\ref{MTZ5}).} \\

\begin{figure}
\centerline{\includegraphics[height=8cm]{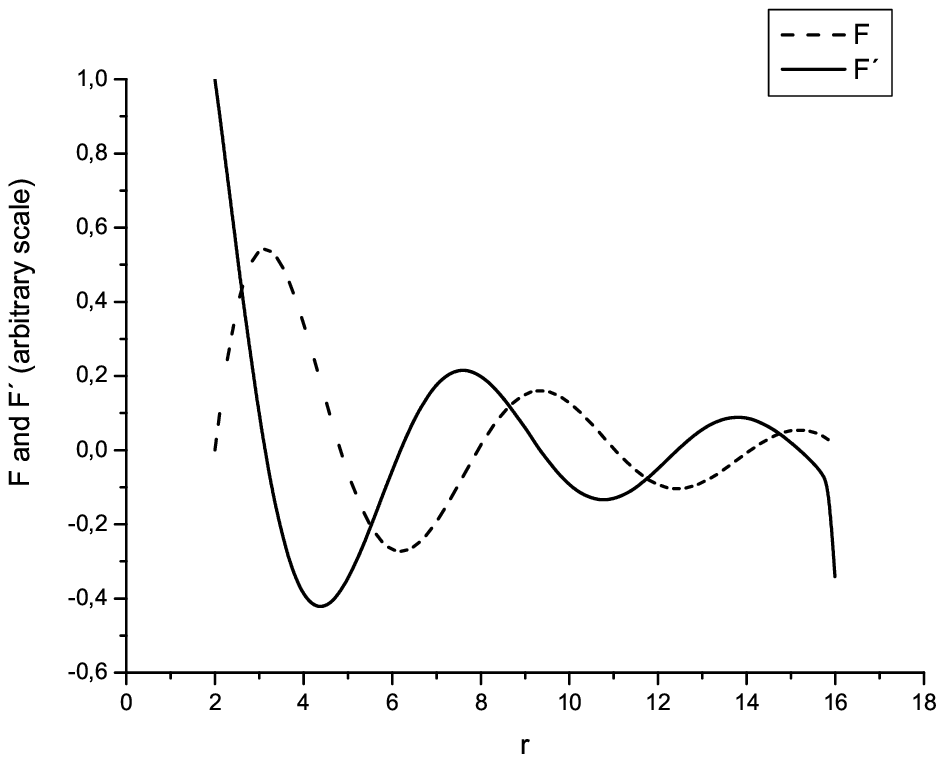}
\includegraphics[height=8cm]{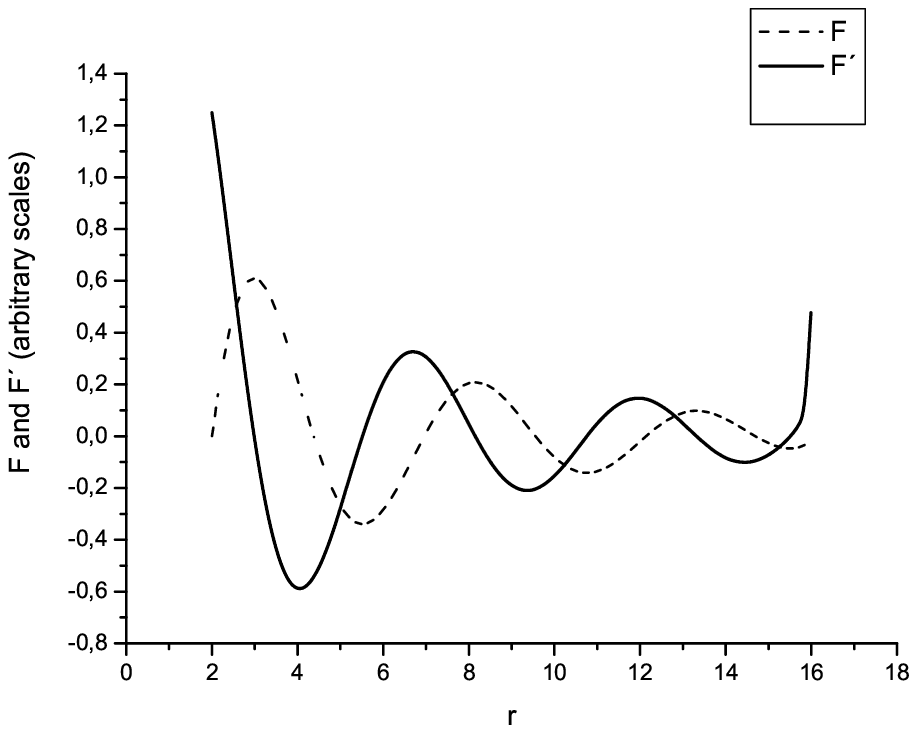}}
\vspace{-1cm} \caption[fins]{ \label{fins} Numerical integration of equation (\ref{ap4}) from $r=r_H$ to $r=r_{\Lambda}$. Equation (\ref{ap5})
guarantees that $F$ will vanish at both horizons, however, $F'$ will generically diverge at $r=r_{\Lambda}$ if it is finite at $r=r_H$, except
for a discrete set of $k$ values. The allowed $k$ values can be spotted by a numerical ``shooting" algorithm. This is illustrated in the figure
above: the left panel shows a numerical integration for $k=1.0$, the right panel a numerical integration for $k=1.2$. It follows that there is
an allowed value $k_o$, with $1.0 < k_o < 1.2$. The integrations were performed for a special theory (\ref{st}), setting
$\Lambda=\frac{48}{4745},Q=\frac{146}{16} \Lambda$, and $M=\frac{1755}{16}$, which gives  $r_H=2$ and $r_{\Lambda}=16$. }
\end{figure}

As far as we know, a linear stability analysis of MTZ1 has not yet been done. In what follows we sketch the construction of some particular
unstable modes for the theory (\ref{st}), of the restricted form,
\begin{eqnarray}
\label{ap3}
  \delta \phi(r,t) &=& 0  \nonumber \\
  \delta g_{rr}(r,t) &=& F(r) \exp(k t) \nonumber \\
\delta g_{tt}(r,t) &=&- A F(r) \exp(k t)
\end{eqnarray}
where $\delta$ indicates the perturbed part, and $A $ is constant.
Unstable modes would result if we find appropriate solutions for the
perturbation equations with $k$ real and positive. Replacing this
Ansatz in Einstein's and the scalar field equation, and keeping
only linear terms in $F$, the only non trivial equation that results
is of the form,
\begin{equation}\label{ap4}
\frac{d^2 F}{dr^2} = \frac{P_1(r)}{r^7 N(r)} \frac{dF}{dr}
+\frac{P_2(r)}{r^{10}N(r)^2} F(r) +\frac{k^2}{A} F(r)
\end{equation}
where $P_1$ and $P_2$ are polynomials in $r$ with coefficients that depend only on $\Lambda$, $M$, $Q$, and $A$, which are therefore regular in
the relevant range in $r$, that is for $r_H \leq r \leq r_{\Lambda}$, with $r=r_H$ (the event horizon), and $r=r_{\Lambda}$ (the cosmological
horizon) corresponding to single zeros of $N$. It can be checked that, the general solution of (\ref{ap4}), near one of the zeros of $N$, which
are the singular points of (\ref{ap4}), is of the form,
\begin{equation}
\label{ap5}
    F(r) \simeq c_1 (r-r_p) + c_2 \sqrt{|r-r_p|}
\end{equation}
where $c_1$, and $c_2$ are arbitrary constants, and $r_p$ is either $r_H$ or $r_{\Lambda}$. This means that the general solution of (\ref{ap4})
vanishes at the horizons, but only those solutions with $c_2=0$ at both $r=r_H$ and $r=r_{\Lambda}$ are acceptable as perturbations, because for
$c_2\neq 0$ the derivatives of $F(r)$ are singular. This implies that appropriate solutions, if they exist, satisfy a boundary value problem,
with $k^2$ the corresponding eigenvalue. Considering now a numerical integration of (\ref{ap4}), there is no difficulty in imposing regularity
for, say, $r= r_H$, but, for general $A$ and $k$ the resulting solution would be singular for $r=r_{\Lambda}$. We notice, however, that for
large enough $k$, and $r$ not close to the horizons, (\ref{ap4}) behaves approximately as,
\begin{equation}\label{ap6}
\frac{d^2 F}{dr^2} \simeq \frac{k^2}{A} F(r).
\end{equation}
Therefore, if we take $A<0$, $F(r)$ will oscillate between positive and negative values in the region $r_H \leq r \leq r_{\Lambda}$. This
implies that, for $A<0$, imposing the condition that $c_2$ vanishes for both $r=r_H$ and $r=r_{\Lambda}$ turns (\ref{ap4}) into a boundary value
problem determining the allowed values of $k$. Note that (\ref{ap5})  guarantees that the perturbation will vanish at both horizons. Clearly,
there is no upper bound on the allowed $k$ values. Therefore, the linear perturbation problem leads to solutions that diverge arbitrarily fast
from MTZ1. Figure \ref{fins}  illustrates a ``shooting" approach to the problem of finding appropriate values for $k$: $Q,M$ and $\Lambda$ were
chosen so that $r_H=2$ and $r_{\Lambda}=16$, and (\ref{ap4}) was numerically integrated from $r=r_H$, setting $c_2=0$ at this horizon (see eq.
(\ref{ap5})). Generically, the solution at $r=r_{\Lambda}$ will also be of the form (\ref{ap5}), but with $c_2 \neq 0$,  then $F'$ will diverge
there. Requiring that $F'$ be finite at {\em both} horizons gives a discrete set of possible $k$ values. The left panel of the figure shows a
numerical integration performed with $k=1.0$, the right panel shows a numerical integration with $k=1.2$. The fact that at both horizons the
behavior is as in (\ref{ap5}) guarantees the vanishing of $F$. It is clear from Figure \ref{fins} and continuity arguments,  that,  for some
value of $k$ in this interval there is a solution with a finite derivative at $r_{\Lambda}$.

The analysis carried out in \cite{HTWY} indicates that the solutions
MTZ2 are also unstable under linear, spherical perturbations. Once
again, if one attempts to solve the linear perturbation equations
for the spherically symmetric mode, one finds solutions that grow in
time arbitrarily fast \cite{HTWY}. This may be traced to the fact
that the perturbation ``potential'' (of the Regge-Wheeler like
equation) is singular for $r=2M$, a rather peculiar situation, since
the metric (\ref{exa1}) and the scalar field are  smooth in the
range between the event and cosmological horizons, in particular at
$r=2M$, since $r_2 < 2M < r_{3}$.\\

More generally, the problem of solving the linearized equations for
arbitrary (i.e., not restricted to spherically symmetric)
perturbations can be approached  by decomposing in angular modes in
the usual way, and projecting onto $S^2$ harmonic vector and scalar
fields, but, as we have checked, this leads to and extremely
intricate set of equations that is difficult to deal with. Notice
however that in order to prove instability, it is certainly
sufficient to exhibit a single unstable mode, as was done above
for MTZ1 and in \cite{HTWY} for MTZ2. \\

An important point is that under  the radial perturbations  above
and in \cite{HTWY}, the perturbed metrics leave the restricted
family  $g_{tt} = -1/g_{rr}$, getting  into the general space of
static and spherically symmetric spacetimes (\ref{met1}). This
suggests that the peculiar behavior of the MTZ solutions under
perturbations may be related to the restricted nature of the space of
solutions of the form  (\ref{MTZ5}). For instance, the perturbation
method might not be applicable because in the general case $N_2 \neq
1/N_1$, there are solutions which are {\em locally} arbitrarily
close the unperturbed one in the family (\ref{met1}), but with very
different {\em global} behaviour. This, as we show in this Paper, is
precisely the case for the MTZ family of solutions.
\\

\section{The Einstein equations} \label{metric}

We generalize the metric Ansatz of \cite{Martinez:2002ru} by considering instead
of (\ref{MTZ5}), a static spherically symmetric metric and scalar field of the form (\ref{met1}):
\begin{equation}\nonumber
ds^2 = -N_2(r) dt^2 + N_1(r) dr^2 + r^2 d\Omega^2, \;\; \phi = \phi(r)
\end{equation}
Recall from the previous Section that (\ref{se}) always hold, and thus (\ref{el-b}) can
be replaced with (\ref{MTZ3.1}). Inserting
(\ref{met1}) in ($N_1$ times) Eq. (\ref{MTZ3.1}) gives:
\begin{equation} \label{efi}
0 = \phi'' + \frac{1}{2}  \phi' \left[  \frac{ N_2{}'}{N_2} -  \frac{ N_1{}'}{N_1} + \frac{4}{r}
\right] - \frac{2}{3}\;  N_1 \; \phi   \; \left[ 6 \alpha \phi^2 + \Lambda \right].
\end{equation}
Also,
${\cal G}_{\mu\nu}:=G_{\mu\nu}+\Lambda
g_{\mu\nu}-T^{\phi}_{\mu\nu}$
is diagonal, with
${\cal G}_{\theta \theta} \propto {\cal G}_{\phi\phi}$
, thus (\ref{el-a}) gives three non trivial equations. The first two are:
\begin{multline} \label{ett}
0 =
\left( \frac{6 r^2\; N_1}{N_2} \right) {\cal G}_{tt} =
   -\frac{{N_1}'}{N_1}
 r \, \left[ (\phi^2-6) + r \phi' \phi \right]
- N_1  \left[ (\phi^2-6) + 6r^2 (\Lambda + \alpha \phi^4) \right] \\ + \left[4 r \phi \phi'  +
(\phi^2 -6)
 + 2 r^2 \phi \phi'' - r^2 \phi'{}^2 \right],
\end{multline}
\begin{multline} \label{err}
0 = 6 r^2 \; {\cal G}_{rr} =  -\frac{{N_2}'}{N_2} r \left[ (\phi^2-6) + r \phi' \phi \right]
  +N_1  \left[ (\phi^2-6) + 6r^2 (\Lambda + \alpha \phi^4) \right]  \\
- \left[ 3 r^2  \phi'^2  + 4 r \phi \phi' + (\phi^2-6) \right],
\end{multline}
The ${\cal G}_{\theta\theta}$ equation is seen (after some
work) to actually follow from (\ref{efi}), (\ref{ett}) and
(\ref{err}), so it will not be needed. The field equations conform a
system of three ODEs, (\ref{efi})-(\ref{err}),
 on three unknown functions
$N_1,N_2$ and $\phi$. \\

It is apparent from (\ref{ett})-(\ref{err}) that the case where $
\left[ (\phi^2-6) + r \phi' \phi \right] \equiv 0$ is special. If
such a solution exists then $\phi=\sqrt{6 r^2 + C}/r$ and equations
(\ref{ett})-(\ref{err}) force $C=0$ (i.e., $\phi=\sqrt{6}$), and
$\alpha=-\Lambda/36$. Under these conditions the remaining field
equation,  eq. (\ref{efi}), is also  satisfied. This is of course
solution MTZ1 in the special case $\alpha=-\Lambda/36$, eq. (\ref{1MTZs}). \\
If, on the other hand, $ \left[ (\phi^2-6) + r \phi' \phi \right]
\not \equiv 0$ (in particular $\phi^2 \not \equiv 6$), we find,
after some work on (\ref{efi})-(\ref{err}) that

\begin{eqnarray}\label{a3}
{\frac {d^{2}\phi}{d{r}^{2}}} &= & {\frac { \left[  \left( 2\,
\left( 9\,\Lambda- \phi^{2}\Lambda+3\,\alpha\,   \phi^{4}  \right)
{r}^{2}+3\, \phi^{2}-18  \right)  \phi'   -2\,r\phi \left(
\phi^{2}-6  \right)  \left( 6\,\alpha\,  \phi^{2
}+\Lambda \right)  \right] { N_1}  }{3 r \left( 6 - \phi  ^{2} \right) }}\nonumber \\
& & -{\frac {        \left( 6- \phi^{2}+  \left(\phi' \right)
^{2}{r}^{2}+2\,r\phi    \phi'    \right) \phi'  }{ r \left( 6 - \phi
^{2} \right) }},
\end{eqnarray}
and also that   we can write $T^{\phi}_{\mu \nu}$  just
in terms of $N_1,N_2,\phi$ and $\phi'$, using  the  orthonormal basis
 $\hat t ^{\mu} = {N_2}^{-1/2} \; \partial_t, \;
\hat r^{\mu} = {N_1}^{-1/2} \; {\partial}_{r},\; \hat \theta ^{\mu} = r^{-1}
 {\partial}_{\theta},\; \hat \phi ^{\mu} = (r \sin(\theta) )
 ^{-1} {\partial}_{\phi}$:
 \begin{equation} \label{tmunu}
 T_{\mu \nu} = \rho \; \hat t_{\mu} \hat t_{\nu} + p_r \; \hat r_{\mu} \hat r_{\nu} +
 p_{\theta} \; \hat \theta_{\mu} \hat \theta_{\nu} + p_{\phi} \; \hat \phi_{\mu} \hat \phi_{\nu}.
 \end{equation}
 Here
 \begin{eqnarray} \nonumber
\rho &=& {\frac { \left( -12\,{\phi}^{5}{r}^{2}\alpha+ \left( \Lambda\,{r}
^{2}-3 \right) {\phi}^{3}+ \left( 18-18\,\Lambda\,{r}^{2} \right) \phi
 \right) {\it \phi'}+6\,\alpha\,{\phi}^{6}r+ \left( -36\,r\alpha+
\Lambda\,r \right) {\phi}^{4}-6\,\Lambda\,r{\phi}^{2}}{3 r \left( -6+{
\phi}^{2} \right)  \left( -6+{\phi}^{2}+r\phi\,{\it \phi'} \right) }}\\ \nonumber
&&+{\frac {6\,\phi\,{r}^{2}{{\it \phi'}}^{3}+ \left( 18\,r+9\,r{\phi}^
{2} \right) {{\it \phi'}}^{2}+ \left( 3\,{\phi}^{3}-18\,\phi \right) {
\it \phi'}}{3 r \left( -6+{\phi}^{2} \right)  \left( -6+{\phi}^{2}+r\phi
\,{\it \phi'} \right) {\it N_1}}}
\\ \nonumber
p_r &=& {\frac { \left( -1+\Lambda\,{r}^{2} \right) \phi\,{\it \phi'}+r{\phi}^{
2} \left( 6\,\alpha\,{\phi}^{2}+\Lambda \right) }{r \left( -6+{\phi}^{
2}+r\phi\,{\it \phi'} \right) }}-3\,{\frac {{\it \phi'}\, \left( r{\it
\phi'}+\phi \right) }{r \left( -6+{\phi}^{2}+r\phi\,{\it \phi'} \right)
{\it N_1}}}
\\
p_{\theta} &=&  p_{\phi} = \frac{\phi^2 (6 \alpha  \phi^2 + \Lambda)}{3 (\phi^2-6)} - \frac{\phi' (r \phi' +
 2 \phi)}{r (\phi^2-6) N_1} \nonumber
\end{eqnarray}

These formulae do not hold, of course, at those isolated points where
 $\left( -6+r  \phi    \phi' +   \phi ^{2} \right)=0$
or  $\phi^2 = 6$. The fact that this system is singular at points
where $\phi^2=6$ is clearly related to the singular nature of the
linearized  equations for perturbations of MTZ2 (\ref{exa1}) at
$r=2M$, where $\phi=\sqrt{6}$ \cite{HTWY}. This is so because the
perturbed metric (eq.(9)) and scalar field (eq. (11)) in \cite{HTWY}
are, in the static case, of the form (\ref{met1}). In principle,
whether a solution is also singular at such a point depends
critically on the behavior of the numerators of equation (\ref{a3}).
The exact solutions (\ref{exa1}) represent cases where this
singularity is canceled, but other possibilities should be expected.\\

For the MTZ1 solution in the special case $\alpha = -\Lambda/36$,
eq. (\ref{1MTZs}), we cannot use (\ref{tmunu}), and the
energy-momentum tensor in this case is given by (\ref{ec1}). For
(\ref{1MTZ}),  using  (\ref{tmunu}) we get the expected result
$T^{\phi}_{\mu \nu} = 0$.

\section{Solutions with a regular horizon} \label{reghor}

In this Section we consider solutions of the field equations that:
(i) contain a regular event horizon at $r=r_0$, with $r_0>0$, and (ii)
satisfy the weak and dominant energy conditions in some open
neighborhood  $r_0 < r < r+\epsilon$ outside the horizon. \\

 The regular horizon condition
implies that there exists a neighborhood of $r=r_0$, where the
functions $N_1$, $N_2$ and $\phi$ admit expansions of the form,
\begin{eqnarray}
\label{hori1}
\phi &=& a_0 +a_1 (r-r_0) +a_2 (r-r_0)^2+\dots \nonumber \\
  N_1 &=& b_{-1}(r-r_0)^{-1} + b_0 +b_1 (r-r_0) + \dots \\
  N_2 &=& c_1 (r-r_0) +c_2 (r-r_0)^2+c_3 (r-r_0)^3+ \dots \nonumber
\end{eqnarray}
where $a_i$, $b_i$, and $c_i$ are constants coefficients. The proper
signature of the metric imposes $b_{-1} > 0$ and $c_1 >0$, although
$c_1$ is otherwise arbitrary because of the freedom of rescaling of
$t$. We also impose $a_0 \geq 0$, making using of the invariance of
the equations under $\phi \to -\phi$.
We will find it convenient to introduce the dimensionless horizon radius
\begin{equation}\label{zo}
z_o := r_o \sqrt{\Lambda}.
\end{equation}
In Section \ref{4b} we arrive at a description of  the subset of the
$(z_o,a_0)$  plane for which conditions (i) and (ii) above are satisfied
in the special case $\alpha=-\Lambda/36$ (see Figure \ref{modspace}).\\

Replacing the expansions
(\ref{hori1}) in equations (\ref{efi})-(\ref{err}) we obtain
relations between the coefficients by equating powers in $r-r_0$.
From the algebraic equations obtained by matching the lowest order
non trivial terms we learn that the ``special" case $\alpha = - \Lambda/ 36$
requires separate treatment.\\

\subsection{Generic theories ($\alpha \neq -\Lambda/36$)}

To lowest order we obtain
\begin{eqnarray} \label{a1}
a_{{1}} &=& \frac {2 r_0 a_{{0}} \left( {a_{{0}}}^{2}-6 \right)  \left( 6
\,\alpha\,{a_{{0}}}^{2}+\Lambda \right) }{3\,{a_{{0}}}^{2}-2\,{a_{{0}
}}^{2}\Lambda\,{r_0}^{2}+6\,\alpha\,{a_{{0}}}^{4}{r_0}^{2}-18+18\,\Lambda
\,{r_0}^{2}} ,\\ \label{bm1}
b_{{-1}} &=& {\frac{-3\,  {r_0} \left( -6+{a_{{0}}}^{2} \right) }{18-3\,{a_{{0}
}}^{2}-18\,\Lambda\,{r_0}^{2}-6\,\alpha\,{a_{{0}}}^{4}{r_0}^{2}+2\,{a_{{0}
}}^{2}{r_0}^{2}\Lambda}}
\end{eqnarray}
This suggest that we study the  cases A) $a_1=0$ and B) $a_1 \neq 0$ separately.\\

\noindent
{\em Case A.1, $a_0=0$:} in this case, by solving iteratively for the higher order terms, assuming
$r_0 >0$, we are led to the Taylor expansion of Schwarzschild - dS  (S-dS) space:
\begin{equation} \label{ds}
\phi(r)=0, \; N_1(r) = \left[ 1 - \frac{r_0 (3-\Lambda {r_0}^2)}{3 r} - \frac{\Lambda r^2}{3}
\right]^{-1}, N_2(r) = \frac{c_1  r_0}{(\Lambda r_0{}^2 - 1) N_1(r)}
\end{equation}
Matching $N_1 = 1 - 2M/r - \Lambda r^2/3 = -\frac{\Lambda}{3r} (r+r_1)(r+r_2)(r+r_3),$ gives
$r_1=-(r_2+r_3)$. We want, say $r_2=r_0$ (event horizon), $r_3=$ cosmological horizon,
then $0 < r_2 < r_3$. This implies $M = \Lambda (r_2+r_3)r_2r_3/6 >0$ and
\begin{equation} \label{ds2}
z_o  < 1.
\end{equation}

\noindent
{\em Case A.2, $a_0=\sqrt{6}$:} this leads to $b_{-1}=0$, see eq. (\ref{bm1}).
Interestingly, no solution
with a regular horizon and $\phi(r_0)=\sqrt{6}$ exists {\em in the generic theory} if
we assume the scalar field admits a Taylor expansion around the horizon.\\

\noindent
{\em Case A.3, $a_0 = \sqrt{\frac{-\Lambda}{6 \alpha}}$:} in this case we  obtain, once again,
$a_j=0$ for $j>0$, i.e., $\phi= \sqrt{\frac{-\Lambda}{6 \alpha}}$, together with
\begin{multline} \nonumber
 N_1(r) = \frac{r_0}{1-\Lambda {r_0}^2} (r-r_0)^{-1} + (1-\Lambda {r_0}^2)^{-2} \\
+ \frac{\Lambda r_0 (4-\Lambda {r_0}^2)}{3 (1-\Lambda {r_0}^2)^{3}} (r-r_0)
+ \frac{r_0 (3 \Lambda {r_0}^2 + 1 - \Lambda^2 {r_0}^4)}{3 (1-\Lambda {r_0}^2)^{4}} (r-r_0)^2
+ \dots
\end{multline}
which is the Taylor expansion around $r=r_0$ of MTZ1, eq.(\ref{1MTZ}),  written as
$$
N_1(r) = \left[ 1 + \frac{r_0 (\Lambda {r_0}^2-3)}{3r}-\frac{\Lambda r^2}{3} \right]^{-1}
$$

\noindent
{\em Case B, $a_1 \neq 0$:} This case   is extremely
 complex to deal with in the general situation.
Since the main motivation of this work is to understand the behavior of the linearized
field equations around the solution MTZ2 (\ref{exa1}) for the theory
 $\alpha = -\frac{\Lambda}{36}$, we restrict our attention to
special theories from now on. \\

\subsection{Special Theories $\alpha=-\Lambda/36$} \label{4b}

To lowest order, for the special theories we obtain

\begin{equation}
a_1 = \frac{- 2 \Lambda r_0 a_0 ({a_0}^2-6)}{18 - {r_0}^2 \Lambda ({a_0}^2 +18)}, \;
b_{-1} = \frac{18 r_0}{18 - {r_0}^2 \Lambda ({a_0}^2 +18)},
\end{equation}

This suggests that we study the cases $a_0 = 0$ and $a_0 = \sqrt{6}$ separately.\\

\noindent
{\em Case A: $a_0=0$:} To no surprise, we are led back to Schwarzschild de Sitter space, eqs. (\ref{ds}) and
(\ref{ds2}).\\

\noindent
{\em Case B: $a_0=\sqrt{6}$:}  The higher order terms of (\ref{efi}), (\ref{ett}) and (\ref{err})
imply give $a_j=0, j > 0$, i.e., {\em any solution with $\phi(r_0)= \sqrt{6}$ must
satisfy $\phi(r) = \sqrt{6}$ for all $r$}.
From the comments  in Section \ref{intro}, we know that the only field equation for the metric
 in this case is  $R = 4 \Lambda$, which reads
\begin{equation} \label{r4l}
- \frac{N_2{}''}{N_1 N_2} + \frac{ ({N_2}')^2}{2 N_1 {N_2}^2} + \frac{{N_1}' {N_2}'}{2
{N_1}^2 N_2} - \frac{2 {N_2}'}{r N_1 N_2} + \frac{ 2 {N_1}'}{r {N_1}^2} +
 \frac{2(N_1 -1)}{r^2 N_1} = 4 \Lambda.
\end{equation}

In principle, this gives us an infinite number of solutions for the
Ansatz (\ref{met1}), since, given, say $N_2$, $R = 4 \Lambda$ is a
first order  ODE for $N_1$. In particular,  given $N_2$ as in
(\ref{hori1}) and any $b_{-1} >0$, the algebraic equations for the
 remaining $b_j 's$ admit a solution. Inserting this solution in the energy
 momentum tensor $T^{\phi}_{\mu \nu} = G_{\mu \nu} + \Lambda g_{\mu \nu}$ and using
 the orthonormal basis in (\ref{tmunu}) gives
 \begin{equation}\label{rho}
\rho = \rho_o  + {\cal{O}}\left((r-r_0)\right),\;\;\; \rho_o := \frac{ b_{-1} (1 - \Lambda {r_0}^2) - r_0}{b_{-1} {r_0}^2} ,
\end{equation}
and
 \begin{equation}\label{pr}
 \frac{p_{\theta}}{\rho} = - \frac{p_r}{\rho} = 1 + \left( \frac{ 2 c_1 b_{-1} (2 \Lambda
{r_0}^2-1)
+ 4 c_1 r_0 + 2 c_2 {r_0}^2}{r_0 c_1 [ r_0 +b_{-1} (\Lambda {r_0}^2-1)]} \right) (r-r_0) +
{\cal{O}}\left((r-r_0)^2\right)
\end{equation}
The condition $\rho_o > 0 $ is equivalent to
\begin{equation}
\frac{1-z_o{}^2}{z_o} > \frac{1}{ \sqrt{\Lambda} b_{-1}} \Leftrightarrow 0 < z_o <
\frac{\sqrt{1 + 4 \Lambda b_{-1}{}^2}-1}{2 \sqrt{\Lambda} b_{-1}},
\end{equation}
and thus $z_o < 1$, as happens for generic theories, eq. (\ref{ds2}). To satisfy the strong energy condition in some open $r$ interval $r_0 < r <
r_0 + \epsilon$ we require that the $(r-r_0)$ coefficient in (\ref{pr}) be negative, and this can always be satisfied by a proper choice of $c_2$,
thus proving that there are
local solutions satisfying the energy conditions right outside a regular horizon.\\

 It is not hard to see
 that, out of the infinitely  many solutions for the ODE (\ref{r4l}),
 the only one satisfying  $N_1 N_2
 \equiv 1$ is MTZ1 (\ref{1MTZs}). Given  MTZ1 with
 positive $\Lambda$ and positive $Q$ (required by
 the energy conditions, see (\ref{ec1})), one can easily see that in order  to avoid naked singularities
 the quartic   polynomial $r^2 N(r)$ has to  have four real roots,  one negative and three positive:
$ -r_4  < 0 < r_1 < r_2 < r_3$, with $r_2=r_0$  the event horizon and $r_3$ the
cosmological horizon. Then matching (\ref{1MTZs}) with
\begin{equation}
N(r) = - \frac{\Lambda}{3 r^2} (r+r_4) (r-r_1) (r-r_2) (r-r_3),
\end{equation}
gives  $r_4 = (r_1+r_2+r_3)$, a positive mass
\begin{equation}
M = \frac{(r_1+r_2)(r_1+r_3)(r_2+r_3)}{2  \sum_{i \leq j \leq 3} r_i r_j},
\end{equation}
and
\begin{equation}
\Lambda = \frac{3}{ \sum_{i \leq j \leq 3} r_i r_j}, \;
Q = \frac{(r_1+r_2+r_3)(r_1 r_2 r_3)}{  \sum_{i \leq j \leq 3} r_i r_j}.
\end{equation}
In particular, given the domain $0 < r_1 < r_2 < r_3 < \infty$, one finds that
 $z_0 = r_2 \sqrt{\Lambda}$
satisfies the constraint (compare to (\ref{ds2}) and (\ref{c2}))
\begin{equation}
0 < z_o  < 1
\end{equation}

 \noindent
{\em Case C: $a_0 \neq 0, \sqrt{6}$:}
In this case  we find
that all the coefficients in
(\ref{hori1}) may be written, e.g., in terms of $r_0$,
and $a_0$. The leading terms are of the form,
\begin{eqnarray}
\label{hori2} \phi &=& a_0 + \frac{2 (a_0^2-6)\Lambda r_0
a_0}{\Lambda r_0^2a_0^2+18 \Lambda r_0^2 -18} (r-r_0) + \frac{4
(a_0^2-6)^2\Lambda^2 r_0^2 a_0}{(\Lambda r_0^2a_0^2+18
\Lambda r_0^2 -18)^2} (r-r_0)^2+\dots  \nonumber \\
N_1 &=& \frac{18 r_0}{(18-18 \Lambda r_0^2 -\Lambda r_0^2 a_0^2)(r-r_0)}
 -\frac{ 36(\Lambda r_0^2a_0^2-9)}
{(18-18 \Lambda r_0^2 -\Lambda r_0^2 a_0^2)^2} + \dots \\
N_2 &=& c_1 \left[ (r-r_0) +\frac{2 (\Lambda r_0^2 a_0^2-9)}{r_0(18-18
\Lambda r_0^2 -\Lambda r_0^2 a_0^2)} (r-r_0)^2+ \dots \right] \nonumber
\end{eqnarray}
where, as already noticed, $c_1 >0$, but it is otherwise arbitrary.
This implies that the condition for the existence of a regular
horizon leads, in general, to
a two-parameter ($r_0$ and
$a_0$) family of solutions. We notice, for reference, that  the exact solution MTZ2
(\ref{exa1}) corresponds to the  one parameter subfamily for which
\begin{equation}
\label{exa2}
  r_0 =  \frac{\sqrt{3}}{2\sqrt{\Lambda}}-\frac{\sqrt{3
  \Lambda-4M\sqrt{3 \Lambda^3}}}{2\Lambda} , \;\;
  a_0 = \frac{\sqrt{6} M}{r_0-M} , \;\;  0 \leq M \leq \frac{1}{4} \sqrt{\frac{3}{\Lambda}}
  \end{equation}
  with $c_1$ chosen as $c_1 = \frac {(18-18 \Lambda r_0^2 -\Lambda r_0^2 a_0^2)}{18 r_0}$
  in (\ref{exa1}). The limit case $M=0$ gives just
de Sitter spacetime with no scalar field.\\

As explained in Section \ref{intro}, MTZ1 and MTZ2 are the only solutions
with $N_1=N_2$. It is important to check  that the expansions
(\ref{hori2}) are consistent with this fact. From (\ref{hori2}) we obtain
\begin{eqnarray}
\label{coso1}
 N_1 N_2 & =&  {\frac {18 r_0\,c_{{1}}}{18-{r_0}^{2} \left( {a_{{0}}}^{
2}+18 \right) \Lambda}}  -  {\frac {24 c_{{1}}\Lambda\,
r_0\,{a_{{0}} }^{2} B }{ \left[ -18+{r_0}^{2} \left(
{a_{{0}}}^{2}+18 \right) \Lambda \right] ^{4}}} \left( r-r_0 \right)
^{2} \nonumber
\\ &  & - {\frac {16 c_{{ 1}}\Lambda {a_{{0}}}^{2} \left[
\Lambda {r_0}^{2} \left( {a_{{0 }}}^{2}-66 \right) +9 \right] B }{
\left[ -18+{r_0}^{2} \left( {a _{{0}}}^{2}+18 \right) \Lambda
\right] ^{5}}} \left( r-r_0
 \right) ^{3}   +{\cal{O}} \left(  \left( r-r_0 \right) ^{4} \right)
  \end{eqnarray}
where,
\begin{equation}\label{B}
    B =
324+{r_0}^{4} \left( {a_{{0}}}^{2}-6 \right) ^{2}{
\Lambda}^{2}-36\,{r_0}^{2} \left( 6+{a_{{0}}}^{2} \right) \Lambda
\end{equation}
Therefore, the condition $N_1(r) N_2(r) =1$ can be imposed only if
$a_0=0$, which is trivial, or if $B=0$. In this case, solving for
$a^2_0$ in terms of the other constants, we find two solutions, but
only one of these leads to acceptable coefficients in (\ref{hori2}).
This solution is given by,
\begin{equation}\label{a01}
a_0^2  = \frac {18+6\,\Lambda\,{r_0}^{2}-12\,\sqrt {3} r_0 \sqrt
{\Lambda}}{\Lambda\,r_0^{2}} = \frac{6 (\sqrt{3}-z_o)^2}{z_o{}^2}
\end{equation}
and it can be checked that this coincides with (\ref{exa2}). \\

Another interesting issue is that of analyzing  the limit $a_0 \to \sqrt{6}$
in (\ref{hori2}). The limit  gives
 $\phi \equiv  \sqrt{6}$, and well defined expansions
for $N_1$, and $N_2$, that can be seen to satisfy the required
 condition  on the metric, $R = 4
\Lambda$. Thus, this is one of the infinitely many  $\phi \equiv
\sqrt{6}$ solutions referred to in Case B above, certainly not MTZ1
(\ref{1MTZs}), since $a_0 \to \sqrt{6}$ in (\ref{coso1}) gives
\begin{equation}\label{N1N2}
N_{{1}}(r)N_{{2}}(r)={\frac {3 r_{{0}}c_{{1}}}{3-4\Lambda
{r_{{0}}}^{2}}}+ {\frac {12 c_{{1}}\Lambda r_{{0}}}{ \left( 4
\Lambda {r_{{0}}} ^{2}-3 \right) ^{3}}} \left( r-r_{{0}} \right)
^{2}+ {\cal{O}} \left( \left( r-r_{{0}} \right) ^{3} \right) \neq \text{ constant.}
\end{equation}

We may obtain important information regarding the physical
acceptability of the solutions (\ref{hori2}) by considering the
behavior of the energy-momentum tensor near the horizon. Imposing
the strong and dominant energy conditions on (\ref{hori2}) places
restrictions on the range of the parameters $(r_0,a_0)$. In the
notation of  equation (\ref{tmunu}),
\begin{equation}\label{tmunu0}
\rho = \frac{1}{18} \Lambda a_0^2 -\frac{2 \Lambda a_0^2}{9 r_0}
(r-r_0) + {\cal{O}}\left((r-r_0)^2\right),
\end{equation}
and
\begin{eqnarray}\label{tmunu1}
\frac{p_r}{\rho} & = &  -1-\frac{ 8 \left[\Lambda r_0^2 a_0^2
-6 (\sqrt{3}+r_0 \sqrt{\Lambda})^2\right]\left[\Lambda r_0^2 a_0^2
-6 (\sqrt{3}-r_0 \sqrt{\Lambda})^2\right]} {3 r_0^2 (\Lambda r_0^2
a_0^2 -18 +18 \Lambda r_0^2)^2}(r-r_0)^2 \nonumber \\ & &
+{\cal{O}}((r-r_0)^3) \nonumber \\
\frac{p_{\theta}}{\rho} & = &  1 +\frac{
4\left[\Lambda r_0^2 a_0^2 -6 (\sqrt{3}+r_0
\sqrt{\Lambda})^2\right]\left[\Lambda r_0^2 a_0^2 -6 (\sqrt{3}-r_0
\sqrt{\Lambda})^2\right]} {3 r_0^2
(\Lambda r_0^2 a_0^2 -18 +18 \Lambda r_0^2)^2}(r-r_0)^2 \nonumber \\
& & +{\cal{O}}((r-r_0)^3)
\end{eqnarray}

Therefore, the solutions satisfy the weak energy condition,
(positive energy density) for all $a_0$, but they violate the
dominant energy condition (absolute value of the stresses not larger
than energy density) in the neighborhood of the horizon unless
$a_0$ and $r_0$ are restricted by the conditions,
\begin{equation}\label{restrica01}
 6 (\sqrt{3}-r_0 \sqrt{\Lambda})^2 \leq \Lambda r_0^2 a_0^2 \leq
 6 (\sqrt{3}+r_0 \sqrt{\Lambda})^2
\end{equation}
At the limits  we have
$p_r / \rho =  -1$, and $p_{\theta}/{\rho} =  1$.
The upper limit is further restricted by the condition,
\begin{equation}\label{restrica02}
  \Lambda r_0^2 a_0^2 <
 18-18 r_0^2 \Lambda
\end{equation}
imposed by the condition $N_1 > 0$. All together this implies,
\begin{equation} \label{restrica03b}
\frac{6 (\sqrt{3}-z_o)^2}{z_o{}^2} \leq {a_0}^2 < \frac{18 (1-{z_o}^2)}{{z_o}^2} ,
\end{equation}
Note from (\ref{exa2}), (\ref{a01}), that MTZ2 (\ref{exa1}) saturates the lower bound above, and that
the allowed interval for ${a_o}^2$ is nonempty only if
\begin{equation} \label{restrica03c}
 z_o < \frac{\sqrt{3}}{2} .
\end{equation}
The restrictions for case C can then be summarized by any of the two equivalent
conditions:
\begin{equation} \label{c1}
  6 <  \frac{6 (\sqrt{3}-z_o)^2}{z_o{}^2}  \leq  {a_0}^2 < \frac{18 (1-{z_o}^2)}{{z_o}^2} ,
  \end{equation}
(the first bound in the chain of inequalities following from (\ref{restrica03c})), or
\begin{equation} \label{c2}
 \frac{\sqrt{3}}{1+ \frac{a_0}{\sqrt{6}}}  \leq  z_0 < \frac{\sqrt{3}}{\sqrt{\frac{{a_0}^2}{6}+3}}
 < \frac{\sqrt{3}}{2}  ,
\end{equation}
(the last bound in the chain of inequalities following from $ a_0{}^2 > 6$). This completes
the discussion of case C. \\

\begin{figure}[h]
\includegraphics[height=8cm]{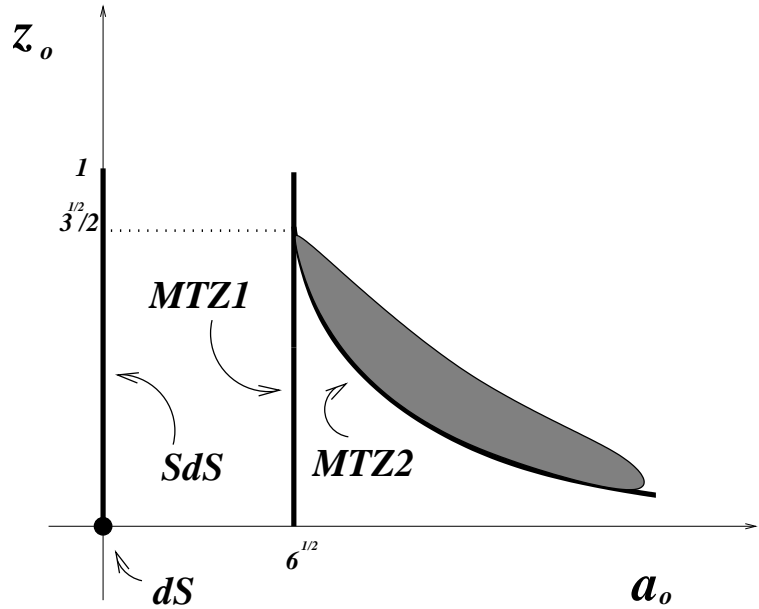}
\caption[modspace]{ \label{modspace} Allowed regions in the
$(z_o,a_o)$ plane for static spherically symmetric local solutions
of the special theory $\alpha=-\Lambda/36$ having a
regular horizon, eqs. (\ref{met1}) (\ref{hori1}) and satisfying the weak and dominant energy condition
in some exterior neighborhood of the horizon.
There is one
solution per point except at the subset $a_0 = \sqrt{6}, \; 0 < z_o
< 1$, where there are (infinitely) many solutions per point. The
horizon radius is $r_0 = \sqrt{\Lambda} \; z_o$ and the value of the
scalar field at the horizon is $\phi(r_0) = a_0$. The $a_0=0, z_o <
1$ solutions are Schwarzschild -  de Sitter,
 the MTZ1 solutions lie on the segment $a_0 = \sqrt{6},
\; 0 < z_o < 1$, the MTZ2 solutions on the lower edge $z_o
= \frac{\sqrt{18}}{\sqrt{6}+ a_0}$ of the shaded $a_0 >
\sqrt{6}$ region. }
\end{figure}

Let us recapitulate on  what we have found by seeking local solutions of the form (\ref{hori1})
for the special theories $\alpha= -\Lambda/36$, satisfying the weak and dominant energy conditions
outside the horizon.  We have used  the $\phi \to -\phi$ symmetry of the field equations
to restrict our considerations to $\phi(r_0) =: a_0 \geq 0$ and found that:
\begin{itemize}
\item If $a_0=0$ then $\phi(r) \equiv 0$ and the metric  is Schwarzschild de
 Sitter. The constraint $z_o < 1$ is required to
assure there is an event horizon hiding the singularity, and an exterior cosmological
horizon.
\item For $0 < a_0 < \sqrt{6}$ there are no solutions satisfying the weak and dominant
energy conditions outside the horizon.
\item If $a_0 = \sqrt{6}$ then $\phi(r) \equiv  \sqrt{6}$, and
 there are infinitely many solutions $(N_1(r),N_2(r))$,
for every $z_o < 1$, (see eq.(\ref{r4l})), some of them satisfying the energy conditions.
\item If $a_0 >\sqrt{6}$ then there is one solution satisfying the desired energy conditions
for every pair $(a_0,z_o)$ satisfying
(\ref{c2})
\end{itemize}
The situation is summarized in Fig \ref{modspace}.\\

The natural question to ask at this point is what is the global behavior of the
local solutions analyzed above. Since solving the system (\ref{efi})-(\ref{err})
analytically is out of consideration, numerical integrations were performed. The results
are gathered in the following Section.

\section{Special theories: Numerical analysis of the $\phi(r_0) > \sqrt{6}$
solutions}\label{numerico}

\begin{figure}[h]
\includegraphics[height=12cm,angle=0]{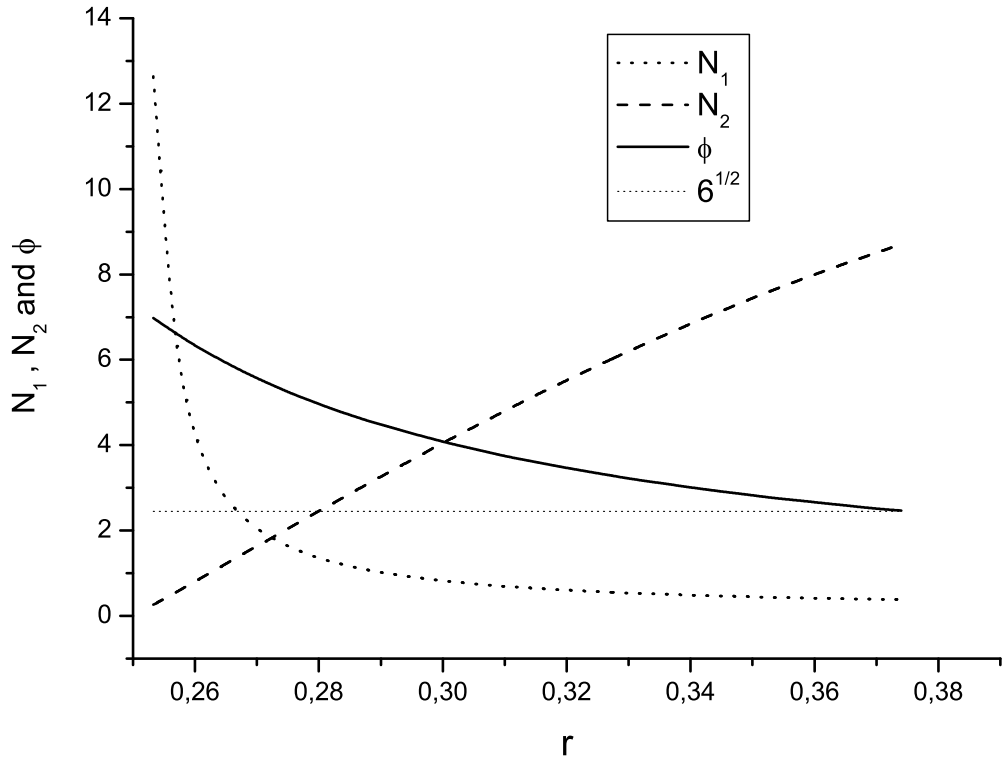}
\caption{\label{MTZnum} {\underline{ Numerically generated MTZ2
solution with $\Lambda=3, r_0=1/4, a_0=3 \sqrt{6}$}}. The vertical
axis displays the correct values of the scalar field, the scales of
$N_1$ and $N_2$ are arbitrary, and were independently chosen to fit
the range of $\phi$ values. Notice the smooth behavior as $\phi$
approaches the regular singularity at $\phi=\sqrt{6}$. }
\end{figure}

The equivalent conditions given in equations (\ref{c1}) and (\ref{c2}) provide a range of values for $r_0$ and $a_0$ such that, locally, the
field equations have a solution with a regular event horizon, with the strong and dominant energy conditions being satisfied right outside the
horizon. The question that naturally arises then is what is the behavior of these solutions  as we move away from $r=r_0$. Since we do not know
of any exact solutions in this range besides the borderline MTZ2, we considered a numerical integration of the system (\ref{efi})-(\ref{err}),
using the expansions (\ref{hori2}) to construct appropriate initial data. A numerical integration requires assigning definite numerical values
to the parameters. We took $c_1=1$, $\Lambda=3$, and $r_0=1/4$, and considered different values of $a_0$ in the interval $3\sqrt{6} \leq {a_0} <
\sqrt{78}$ (eq. (\ref{c1})). To check  the accuracy of the numerical procedure \cite{numerical}, we analyzed as a first example the MTZ2 data
$a_0= 3\sqrt{6}$, which corresponds to $M=3/16$ in eq. (\ref{exa1}). With this choice of $a_0$ we have $\phi^2=6$ for $r=3/8$, and the equations
are formally singular, because of vanishing denominators, for this value of $r$, with the result that the numerical integration stops at that
point. Nevertheless, the numerical solution is well behaved for any $r$ close to but smaller than $3/8$, in correspondence with the regularity
of the exact solution, with a five digit  agreement between the exact and numerical solutions in the plotted range, Figure \ref{MTZnum}.

\begin{figure}[h]
\includegraphics[height=12cm,angle=0]{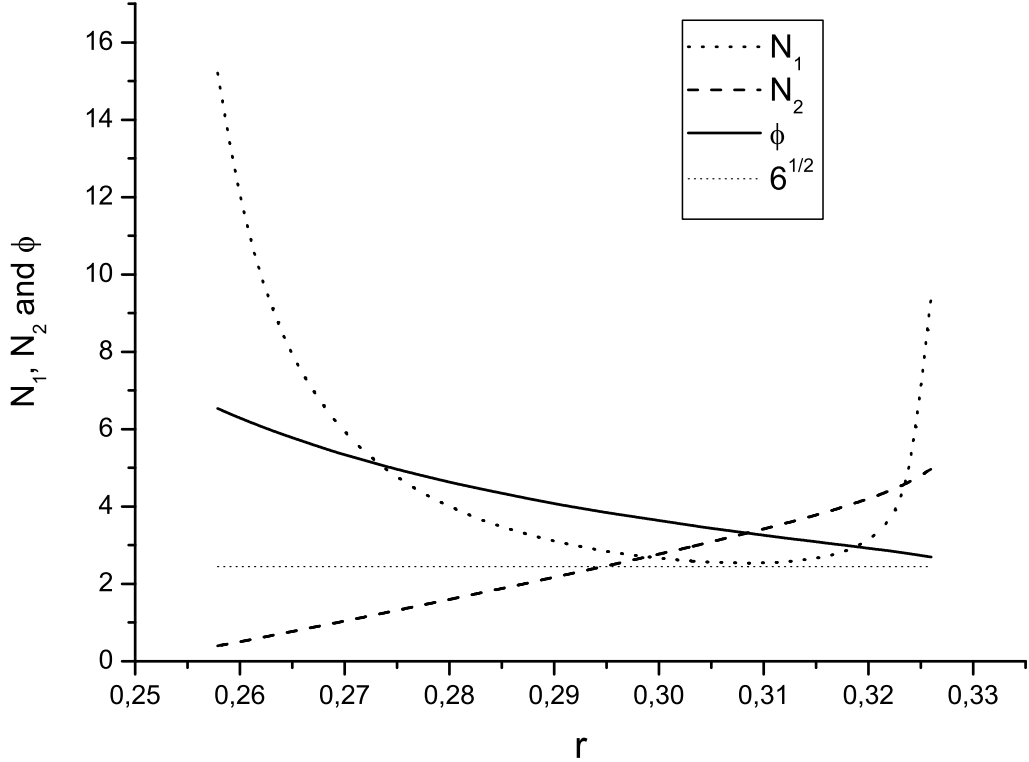}
\caption{\label{aboveMTZ}{\underline{Numerically generated solution
with $\Lambda=3, r_0=1/4,a_0=3 \sqrt{6}+0.3$}}. The vertical axis
displays the correct values of the scalar field, while the scales
for $N_1$ and $N_2$ are arbitrary, and were independently chosen to
fit the range of $\phi$ values. As $r \to r_S$, $\phi$ approaches a
critical value $\phi_c > \sqrt{6}$ and $N_1$ diverges. The critical
value $\phi_c \simeq \sqrt{6}$ in this example because $r_0$, and
$a_0$ are close to  the MTZ2 values.}
\end{figure}

Next we considered, for the same value of  $r_0$,
a number of different allowed values of $a_0$ larger than $3\sqrt{6}$.
The general behavior turned out to be
qualitatively the same in all cases: the numerical
integration shows a singular behavior in $N_1$, as $r$ approaches a
critical value $r=r_S$, while $N_2$ and $\phi$ approach finite
limits, with $\phi \to \phi_c
> \sqrt{6}$. This is illustrated in Figure \ref{aboveMTZ}. It is also found
numerically that $T_t{}^t$, as well as other invariants, approach a
finite limit as $N_1$ diverges. This raises the possibility that the
singular behaviour for $r=r_S$ is only a coordinate effect. In the
next Section we show that this is effectively the case.

\subsection{A coordinate singularity and extensions of the solutions}
\label{coord1}

A detailed numerical analysis of the behavior of $N_1$, $N_2$, and
$\phi$ near the singular point $r=r_S$ indicates that, in general,
we have, for $r < r_S$, and $r \simeq r_S$
\begin{eqnarray}
\label{singb1}
  N_1(r) & \simeq &\frac{ B_1}{r_S-r} \nonumber \\
   N_2(r) & \simeq & C_0 + C_1 \sqrt{r_S-r}  \nonumber \\
  \phi & \simeq & A_0 + A_1 \sqrt{r_S-r}
\end{eqnarray}
where, $A_0,A_1,B_1,C_0,C_1$ are constants that depend on the
solution, and $A_0 > \sqrt{6}$. This suggests the introduction of a
new coordinate $R$, defined by,
\begin{equation}\label{rho1}
R =  \sqrt{r_S-r}
\end{equation}

For this new coordinate the metric takes the form,
\begin{equation}\label{rho2}
ds^2 = -\tilde{N}_2 dt^2 + \tilde{N}_1 d R^2
+(r_S- R^2)^2 d \Omega^2
\end{equation}
where
\begin{eqnarray}
\label{singb2}
  \tilde{N}_1 & = & 4 R^2 N_1 \nonumber \\
   \tilde{N}_2 & = &  N_2 .
\end{eqnarray}

The resulting Einstein and scalar field equations in the new
coordinate $R$ are rather long, and we do not display them here.
We find that, just as in the case of the $r$ coordinate, they are
equivalent to a set of three equations for $\tilde{N}_1$,
$\tilde{N}_2$, and $\tilde{\phi}$. The system has singular
coefficients for $R=0$, but admits regular solutions in the
neighborhood of $R=0$, with $\tilde{N}_1$, $\tilde{N}_2$, and
${\phi}$ having expansions of the form,
\begin{eqnarray}
\label{singb3}
  \tilde{N}_1 & = & \tilde{B}_0 +\tilde{B}_1 R + \tilde{B}_2 R^2+ \dots \nonumber \\
   \tilde{N}_2 & = & \tilde{C}_0 +\tilde{C}_1 R + \tilde{C}_2 R^2+ \dots   \nonumber \\
  {\phi} & =  & \tilde{A}_0 +\tilde{A}_1 R + \tilde{A}_2 R^2+ \dots
\end{eqnarray}
where $\tilde{A}_i$, $\tilde{B}_i$, and $\tilde{C}_i$ are constants,
and $ \tilde{B}_0=4 B_1,  \tilde{C}_0 = C_0, \tilde{C}_1 =C_1$,
in agreement with (\ref{singb1}). Since the transformation
(\ref{rho1}) is defined only for $R >0$ while (\ref{rho2}) is defined also for $R<0$,
the coordinate change (\ref{rho1}) provides a smooth extension of
the original metric (\ref{met1}) through the singular point $r
=r_S$.

We are again here confronted with the lack of explicit exact
solutions, and, therefore, we must resort to a numerical integration
to obtain information on the properties of these solutions. This is
considered in the next Section.

\subsection{Numerical analysis of the continued metrics}
\label{numerico1}

 Given the form (\ref{rho2}) for the
metric, a regular horizon at $R = R_H = \sqrt{r_S-r_0}$ would be characterized
by the functions $\tilde{N}_1$, $\tilde{N}_2$, and ${\phi}$
admitting expansions
\begin{eqnarray}
\label{singb4}
  \tilde{N}_1 & = & \frac{\tilde{b}_{-1}}{R_H-R} +\tilde{b}_0  + \dots \nonumber \\
   \tilde{N}_2 & = & \tilde{c}_1 (R_H-R) +\tilde{c}_2 (R_H-R)^2  + \dots   \nonumber \\
  {\phi} & =  & \tilde{a}_0 +\tilde{a}_1 (R_H-R) + \dots
\end{eqnarray}
where $\tilde{a}_i$, $\tilde{b}_i$, and $\tilde{c}_i$ are constants.
Given a particular solution for (\ref{met1}), with a regular horizon
characterized by given values of $r_0$, and $a_0$, and the remaining
coefficients given by (\ref{hori2}), for which the singularity
appears at $r=r_S$, we have the following relations for the coefficients of
the leading terms:
\begin{eqnarray}
\label{coeffs1}
  \tilde{a}_0 &=& a_0 \nonumber \\
   \tilde{c}_1 &=& 2 c_1 \sqrt{r_S-r_0}  \\ \nonumber
  \tilde{b}_{-1}& = & {\frac {36 r_0 \sqrt {{  r_s}-{  r_0}}}{18- \left( {
a_{{0}}}^{2}+18 \right) {{  r_0}}^{2}\Lambda}}
\end{eqnarray}

We  use  (\ref{coeffs1}) as initial data for a numerical integration
of the equations for $\tilde{N}_1$, $\tilde{N}_2$, and ${\phi}$ in
the region $R_H > R >0$. The numerical integration stops for $R=0$,
but shows that $\tilde{N}_1$, $\tilde{N}_2$, and $\tilde{\phi}$
display a regular behavior arbitrarily close to $R=0$, and allows to
extract the leading coefficients $\tilde{A}_i$, $\tilde{B}_i$, and
$\tilde{C}_i$ in (\ref{singb3}) to compute initial data for the
numerical integration of
the equations in the region $R < 0$, i.e., beyond $r_S$.\\

The main result is that in this extension we find that $\phi
\to \sqrt{6}$, while $\tilde{N_1} \to 0$, and $\tilde{N}_2 \to
\infty$ as $R \to -\sqrt{r_S}$. The energy density  $\rho$,
is found to diverge  as $R \to
-\sqrt{r_S}$.

This situation may be analyzed in general by noticing that $R \to
-\sqrt{r_S}$ corresponds to $r \to 0^+$ in (\ref{met1}) if we change
variables to  $r= r_S -R^2$, so that $r \to 0$ as $R \to
-\sqrt{r_S}$. The numerical results suggest that $(\phi^2-6)\to 0 $,
and $N_1 \to 0$ as some power of $r$, while both $N_2$ and $\rho$
diverge. The detailed behavior near the singularity depends,
however, on some rather delicate cancelations of diverging terms,
and, up to the accuracy achieved so far, we can only draw
qualitative conclusions out of the numerical results \cite{Rscalar}.
To this extent, it appears that the extensions end at a (naked)
singularity, and that the solutions cannot be further extended. This
would imply that the only solution with $a_0 > \sqrt{6}$ containing
a region limited by event and cosmological horizons, where the
energy momentum tensor is compatible with the weak and dominant
energy
conditions, is the exact solution MTZ2 found in \cite{Martinez:2002ru}. \\

Nevertheless, for the problem of understanding the instability found
in \cite{HTWY} we need to study the neighborhood of the MTZ2 curve
in Fig \ref{modspace}, and this includes the dominant energy
violating cases where $6 (\sqrt{3}-r_0 \sqrt{\Lambda})^2 \geq
\Lambda r_0^2 a_0^2$. These are considered in the next Section.

\subsection{ Solutions violating the dominant energy condition } \label{eviol}

A numerical analysis of solutions with  $ 6 (\sqrt{3}-z_o)^2/z_o{}^2 >  {a_0}^2$, -i.e., violating the dominant energy condition near the
horizon and thus (\ref{c1})- reveals  a smooth  behavior of the metric for $r > r_0$, with $\phi \to \sqrt{6}$, and $N_1 \to 0^+$, $N_2 \to 0^+$
as $r$ increases past some value larger than $r_0$. The most remarkable feature of these solutions is that the energy density $\rho(r)$
decreases from its value at the horizon $r=r_0$, changing sign at some $r_1 > r_0$, with $\rho$ taking larger and larger negative values as $r$
increases. As already mentioned, the numerical integration breaks down for sufficiently large values of $r$, but not before the divergence of
$|\rho|$ is clearly established, leading to the conclusion that solutions outside the allowed regions shown in figure \ref{modspace}  contain
features
that make them physically unacceptable.\\

It is interesting that when $a_0$ is slightly smaller than the lower bound forced  in (\ref{c1}), which, as we said, corresponds to the MTZ2
solution (\ref{exa1}), the numerically generated solution remains close to the MTZ2 solution for $r \simeq r_0$, and then they depart completely
from each other as we move away from the horizon. This is a coordinate independent statement, since it is exhibited, e.g., by a  qualitatively
different behavior of the energy density in both cases.

\section{Summary and conclusions} \label{conclus}

We have studied  the theory (\ref{MTZ1}) with $F_{\mu \nu}=0, \Lambda > 0,$
and $\alpha = -\Lambda/36$, and arrived at a  comprehensive
understanding of the space ${\cal M}$ of static, spherically symmetric local solutions
with a regular horizon that satisfy the strong and dominant energy conditions in an
open set bounded by  the horizon. The diagram
in Figure \ref{modspace} shows the $(a_0,z_o)$ plane ($z_o := \sqrt{\Lambda} r_0$,
$a_0 = \phi(r_0)$,
  $r_0$  the horizon radius.) We have proved that there is a one to one, onto correspondence
between the set of $a_0 \neq \sqrt{6}$ solutions in   ${\cal M}$,
and the ($a_0 \neq \sqrt{6}$) shaded region of this plane.
Among these, the only known exact solutions
are MTZ2, eq. (\ref{exa1}) and $\phi \equiv 0$ Schwarzschild - de Sitter spacetime.
The case $a_0 = \sqrt{6}$ is rather peculiar, for every point
in the segment $a_0=\sqrt{6}$,  $0<z_o < 1$ there is not just one, but
 infinitely many local solutions of the field equations admitting a regular horizon, some
 of them satisfying the weak and dominant energy conditions. To this  set belongs
 the other known exact solutions, MTZ1 given in eq. (\ref{1MTZs}).\\

Numerical integrations of the field equations away from the horizon, indicate that
those solutions in the $a_0 > \sqrt{6}$ shaded area are not physically relevant,
since they develop a singularity with infinite energy density,
not protected by a horizon. It
 is rather interesting that, between this singularity and the horizon,
a coordinate singularity was numerically spotted, and appropriate new coordinates
could be  constructed to cross over it. The spheres of symmetry
(i.e., the orbits of the $SO(3)$ isometry group) have a radius (square root
of $(4 \pi)^{-1}$ times their area) that grows from the horizon radius $r_0$
up to a maximum value $r_S$ (where the coordinate change is required),
and then collapses to zero as we approach the above mentioned (spacelike) naked singularity.\\

The unshaded lower region in the $a_0 > \sqrt{6}$ portion of the
$(a_0,z_o)$ plane corresponds to uninteresting solutions of the
field equations. They not only violate the strong energy condition
near the horizon, but also have an energy density $\rho$ that, as we
move away the horizon, becomes negative, and apparently unbounded as
$r$ increases.
(no coordinate change is needed for these solutions). \\

One of the main purposes of the present work was to obtain an
understanding for the extreme instability under perturbations found
in \cite{HTWY} for the metric (\ref{exa1}). From a simple
perspective, given the family of solutions (\ref{exa1}), one would
expect that under a sufficiently small perturbation the system would
radiate some gravitational and scalar field energy, partly to each
horizon, and eventually settle to a static solution of the type
(\ref{exa1}), perhaps with different values of $r_0$ and $a_0$ (or
$M$ in the notation of \cite{Martinez:2002ru}), and therefore, the
results of \cite{HTWY} appear as difficult to understand. The
present analysis, however, indicates that the parameter space for
the static spherically symmetric solutions of the MTZ system indeed
presents a sharp discontinuity at the exact solution, with
neighbouring solutions displaying properties that depart completely
from those of the solution (\ref{exa1}). In particular, the analysis
of Section \ref{coord1} shows that the coordinate system used both
in \cite{Martinez:2002ru} and \cite{HTWY} is inadequate for the
perturbative study, because of the coordinate singularity intrinsic
to that system. But the same analysis shows that even if the
coordinate singularity is avoided, there are solutions that approach
arbitrarily close to (\ref{exa1}) near the black hole event horizon
at $r=r_0$, but then depart from each other with totally different
geometrical properties. In fact, in accordance with (\ref{a01}), for
a given $\Lambda$, the MTZ2 solution is obtained only if $a_0$,
$r_0$, and $\Lambda$ are ``fine tuned'' so that (\ref{a01}) is
satisfied, and any departure from that relation leads either to
solutions with a divergent behaviour for finite $r$ (before a
cosmological horizon is reached), or to solutions with no
cosmological horizon, but with a divergent behaviour for the energy
density.

The final conclusion of our analysis is that there appear to be no
physically acceptable stable solutions of the MTZ system that can be
interpreted as black holes with a cosmological horizon in the
exterior of is event horizon.

\section*{Acknowledgments}

This work was supported in part by grants from CONICET (Argentina) and Universidad Nacional de C\'ordoba.  RJG and GD are supported by CONICET.
This work was also partially funded by FONDECYT (Chile) grants 1051064, 1051056, 1061291, 1071125. The Centro de Estudios Cient\'{\i}ficos
(CECS) is funded by the Chilean Government through the Millennium Science Initiative and the Centers of Excellence Base Financing Program of
Conicyt. CECS is also supported by a group of private companies which at present includes Antofagasta Minerals, Arauco, Empresas CMPC, Indura,
Naviera Ultragas and Telef\'onica del Sur. CIN is funded by Conicyt and the Gobierno Regional de Los R\'{\i}os.  We thank the referees for
comments that led to an improved presentation.


\begin{thebibliography}{99}

\bibitem{Ruffini-Wheeler} R. Ruffini and J. Wheeler, Phys. Today \textbf{24}%
(1), 30 (1971).

\bibitem{Bekenstein:1971hc} J.~D.~Bekenstein,
Phys.\ Rev.\ D\textbf{5}, 1239 (1972).

\bibitem{Teitelboim-No-hair} C.~Teitelboim,
Phys.\ Rev.\ D \textbf{5}, 2941 (1972).

\bibitem{recentwork}  D.~Sudarsky, Class.\ Quant.\ Grav.\ \textbf{12}, 579.
(1995);  A.~Saa, J.\ Math.\ Phys.\  {\bf 37}, 2346 (1996);
A.~E.~Mayo and J.~D.~Bekenstein, Phys.\ Rev.\ D {\bf 54}, 5059
(1996); A.~Saa, Phys.\ Rev.\ D {\bf 53}, 7377 (1996);
J.~D.~Bekenstein, \emph{``Black hole hair: Twenty-five years
after''}, arXiv:gr-qc/9605059; D.~Sudarsky and T.~Zannias, Phys.\
Rev.\ \textbf{D58}, 087502 (1998); K.~A.~Bronnikov and G.~N.~Shikin,
Grav.\ Cosmol.\  {\bf 8}, 107 (2002); D.~Sudarsky and
J.~A.~Gonz\'alez, Phys.\ Rev.\ \textbf{D67}, 024038 (2003);
E.~Ay\'on-Beato, Class.\ Quant.\ Grav.\ \textbf{19}, 5465 (2002);
U.~Nucamendi and M.~Salgado, Phys.\ Rev.\ \textbf{D68}, 044026
(2003); T.~Hertog, Phys.\ Rev.\  D {\bf 74}, 084008 (2006).

\bibitem{Martinez:1996gn}
C.~Mart\'{i}nez and J.~Zanelli, Phys.\ Rev.\  D {\bf 54}, 3830
(1996).

\bibitem{Henneaux:2002wm}
M.~Henneaux, C.~Mart\'{i}nez, R.~Troncoso and J.~Zanelli, Phys.\
Rev.\  D {\bf 65}, 104007 (2002).

\bibitem{BBMB}  N. Bocharova, K. Bronnikov and V. Melnikov, Vestn. Mosk.
Univ. Fiz. Astron. \textbf{6}, 706 (1970). J.~D.~Bekenstein, Annals
Phys.\ \textbf{82}, 535 (1974); Annals Phys.\ \textbf{91}, 75
(1975).



\bibitem{Martinez:2002ru}  C.~Mart\'{i}nez, R.~Troncoso and J.~Zanelli,
Phys.\ Rev.\ \textbf{D67}, 024008 (2003).

\bibitem{Martinez:2004nb}
C.~Mart\'{i}nez, R.~Troncoso and J.~Zanelli, Phys.\ Rev.\ D {\bf
70}, 084035 (2004).

\bibitem{Martinez:2005di}
C.~Mart\'{i}nez, J.~P.~Staforelli and R.~Troncoso, Phys.\ Rev.\  D
{\bf 74}, 044028 (2006).

\bibitem{Martinez:2006an}
C.~Mart\'{i}nez and R.~Troncoso, Phys.\ Rev.\  D {\bf 74}, 064007
(2006).

\bibitem{Torii:2001pg} T.~Torii, K.~Maeda and M.~Narita, Phys.\ Rev.\
\textbf{D64}, 044007 (2001).

\bibitem{Winstanley:2002jt} E.~Winstanley, Found.\ Phys.\ \textbf{33}, 111
(2003).

\bibitem{Winstanley:2005fu}
E.~Winstanley, Class.\ Quant.\ Grav.\  {\bf 22}, 2233 (2005).

\bibitem{Radu:2005bp}
E.~Radu and E.~Winstanley, Phys.\ Rev.\ D {\bf 72}, 024017 (2005).

\bibitem{Hertog:2004bb}
T.~Hertog and K.~Maeda, Phys.\ Rev.\  D {\bf 71}, 024001 (2005).

\bibitem{Hertog:2004dr}
T.~Hertog and K.~Maeda, JHEP {\bf 0407}, 051 (2004).

\bibitem{Zloshchastiev:2004ny} K.~G.~Zloshchastiev,
Phys.\ Rev.\ Lett.\ \textbf{94}, 121101 (2005).

\bibitem{several}
M.~Henneaux, C.~Mart\'{\i}nez, R.~Troncoso and J.~Zanelli, Phys.\
Rev.\ D \textbf{70}, 044034 (2004); M.~Henneaux, C.~Mart\'{\i}nez,
R.~Troncoso and J.~Zanelli, Annals Phys.\  {\bf 322}, 824 (2007);
A.~J.~Amsel and D.~Marolf, Phys.\ Rev.\  D {\bf 74}, 064006 (2006)
  [Erratum-ibid.\  D {\bf 75}, 029901 (2007)];
M.~Natsuume, T.~Okamura and M.~Sato, Phys.\ Rev.\ D \textbf{61},
104005 (2000);
E.~Ay\'{o}n-Beato, A.~Garc\'{\i}a, A.~Mac\'{\i}as and J.~M.~P\'{e}rez-S\'{a}%
nchez,  Phys.\ Lett.\ B \textbf{495}, 164 (2000); G.~Barnich,
``Conserved charges in gravitational theories: Contribution from
scalar fields", arXiv:gr-qc/0211031; J.~Gegenberg, C.~Mart\'{\i}nez
and R.~Troncoso, Phys.\ Rev.\ D \textbf{67}, 084007 (2003);
E.~Ay\'{o}n-Beato, C.~Mart\'{\i}nez and J.~Zanelli, Gen.\ Rel.\
Grav.\ \textbf{38}, 145 (2006).  E.~Ay\'{o}n-Beato,
C.~Mart\'{\i}nez, R.~Troncoso and J.~Zanelli, Phys.\ Rev.\ D
\textbf{71}, 104037 (2005);  M.~Hortacsu, H.~T.~Ozcelik and
B.~Yapiskan, Gen.\ Rel.\ Grav.\ \textbf{35}, 1209 (2003);
 M.~I.~Park,
Phys.\ Lett.\  B {\bf 597}, 237 (2004); M. Ba\~{n}ados and S.
Theisen, Phys.\ Rev.\ D \textbf{72}, 064019 (2005); M.~Salgado,
Class.\ Quant.\ Grav.\ \textbf{20}, 4551 (2003);  A.~Ashtekar,
A.~Corichi and D.~Sudarsky, Class.\ Quant.\ Grav.\ \textbf{20}, 3413
(2003);  A.~Ashtekar and A.~Corichi, Class.\ Quant.\ Grav.\
\textbf{20}, 4473 (2003);  T.~Hertog and S.~Hollands, Class.\
Quant.\ Grav.\  {\bf 22}, 5323 (2005); P.~L.~McFadden and
N.~G.~Turok, Phys.\ Rev.\ D \textbf{71}, 086004 (2005); A.~Biswas
and S.~Mukherji, JCAP \textbf{0602}, 002 (2006);  I.~Papadimitriou,
JHEP {\bf 0702}, 008 (2007); G.~Koutsoumbas, S.~Musiri,
E.~Papantonopoulos and G.~Siopsis, JHEP {\bf 0610}, 006 (2006);
S.~de Haro, I.~Papadimitriou and A.~C.~Petkou, Phys.\ Rev.\ Lett.\
{\bf 98}, 231601 (2007); A.~M.~Barlow, D.~Doherty and E.~Winstanley,
Phys.\ Rev.\ D {\bf 72}, 024008 (2005).

\bibitem{HTWY}
T.~J.~T.~Harper, P.~A.~Thomas, E.~Winstanley and P.~M.~Young, Phys.\
Rev.\  D {\bf 70}, 064023 (2004).

\bibitem{qc}
  B.~L.~Hu and L.~Parker,
  Phys.\ Rev.\  D {\bf 17}, 933 (1978)
  [Erratum-ibid.\  D {\bf 17}, 3292 (1978)]; P.~R.~Anderson,
  Phys.\ Rev.\  D {\bf 32}, 1302 (1985);  N.~G.~Phillips and B.~L.~Hu,
  Phys.\ Rev.\  D {\bf 55}, 6123 (1997)
  [arXiv:gr-qc/9611012];  B.~L.~Hu and Y.~h.~Zhang,
  Phys.\ Rev.\  D {\bf 37}, 2151 (1988);  M.~V.~Fischetti, J.~B.~Hartle and B.~L.~Hu,
  Phys.\ Rev.\  D {\bf 20}, 1757 (1979);  J.~B.~Hartle and B.~L.~Hu,
  Phys.\ Rev.\  D {\bf 20}, 1772 (1979);  J.~B.~Hartle and B.~L.~Hu,
  Phys.\ Rev.\  D {\bf 21}, 2756 (1980).
  \bibitem{wp} D. Noakes, J. Math. Phys. {\bf 24}, 1846 (1983).

\bibitem{Sonego}
  S.~Sonego and V.~Faraoni,
  Class.\ Quant.\ Grav.\  {\bf 10}, 1185 (1993).

\bibitem{numerical} We used a fourth order Runge-Kutta method for
the numerical integrations mentioned in the text.

\bibitem{Rscalar} It turns out that an independent check on the
accuracy of the numerical integration is given by the (numerical)
value of the Ricci scalar $R$ computed using the the numerical
values of $N_1$ and $N_2$. It was found that, in general, this value
remains very close to $4 \Lambda$ (with fractional error less than
$10^{-6}$) as long as neither $N_1$ nor $N_2$ is too large or too
small. We have disregarded any numerical results that do not satisfy
the criteria that this fractional error is less than $10^{-4}$.


\end{thebibliography}
\end{document}